\documentstyle[aps,prb]{revtex}
\begin{document}
\title{Validity of the Gor'kov expansion near the upper critical field in type-II superconductors}
\author{G M Bruun  and V Nikos Nicopoulos}
\address{Department of Physics,
Clarendon Laboratory,
University of Oxford,
Oxford OX1 3PU}
\date{\today}
\maketitle

\begin{abstract}

We have examined the validity of the Gor'kov expansion in the strength of the order parameter 
of type II superconductors near the upper critical field. Although the  degeneracy of the electron levels in a 
magnetic field gives non-perturbative terms in the solution to the Bogoliubov-de Gennes 
equations we find,  contrary to recent claims, that these non-perturbative terms cancel in the expression 
for the thermodynamic potential and that the traditional Gor'kov theory is correct sufficiently close to 
$H_c2$ at finite temperature. We have derived conditions for the validity of the Gor'kov theory which 
essentially state  that the change in the quasiparticle energies as compared to the normal state energies  
cannot be too large compared to the temperature. 
\end{abstract}
\

PACS numbers: 74.20, 74.60 -w

\section{Introduction}
Recently the view has been advocated that the Gor'kov expansion describing type II 
superconductors close to the upper critical field $H_{c2}$ may be invalid~\cite{Bahcall}.
 Even within the mean-field approximation 
there is a possibility of non-perturbative effects arising from the degeneracy  of the Landau levels for 
electrons moving in a magnetic field.  This degeneracy means that even a small perturbation (e.g. 
superconducting order) can change the quasiparticle levels significantly as compared to the normal 
state. This effect has been proposed as a mechanism for the breakdown of the standard perturbation theory
 describing 
type II superconductors even close to $H_{c2}$. This non-perturbative effect should give rise to effects
 such as tails of residual superconductivity above the usual $H_{c2}$, the possibility of the superconducting 
transition being first order, and unusual behavior of the heat capacity, magnetisation etc.\ close to the 
phase boundary~\cite{Bahcall}. In this work we have examined this possibility. We show that for $T=0$ 
there are indeed  terms not contained in the Gor'kov expansion for the difference between the ground state
 energy of the 
mixed state and the normal state. This is in agreement with the results obtained by Bahcall~\cite{Bahcall}
 and is of no surprise since the Gor'kov expansion is essentially a high temperature series. However,
 for $T\neq 0$ we show that the non-perturbative terms in the difference $\Omega_S-\Omega_N$ 
between the thermodynamic potential in the 
mixed state and  the normal state vanish and indeed that the Gor'kov expansion is a convergent series for the 
superconducting order $\Delta({\mathbf{r}})$ not too large. We thus prove incorrect the claim of Bahcall that 
there is a  non-perturbative third order term in the expression 
for $\Omega_S-\Omega_N$. We have derived some criteria for the convergence radius of the Gor'kov 
expansion in the 
order parameter. A comparison between the results of  the Gor'kov expansion and a numerical solution of 
the corressponding Bogoliubov-de Gennes (BdG) equations~\cite{de Gennes} confirms our conclusions. 
 In this paper we work in two dimensions. We do not consider any 
fluctuation effects relevant to high $T_c$ superconductors~\cite{Blatter}.

\section{Mean field theory}
Within the mean-field approximation the mixed state of a type II superconductor is described by the 
solutions to the BdG-equations. In a constant magnetic field the order parameter forms an Abrikosov 
vortex lattice and it is convenient to use a set of single particle states $\phi_{N,{\mathbf{k}}}$ characterized
 by the Landau 
level index $N$ and a wavevector ${\mathbf{k}}$ in the Brillouin zone of the vortex lattice. In this 
basis the BdG-equations split up into a $2N \times 2N$ secular matrix equation for each ${\mathbf{k}}$, 
where $N$ is the number of Landau levels participating in the pairing.  In this basis the BdG 
equations are \cite{Big Mac1}:
\begin{eqnarray}\label{BdG} (\xi_N-E_{\mathbf{k}}^{\eta})u_{N\mathbf{k}}^{\eta}+\sum_{M}F_{
{\mathbf{k}}NM} v_{M\mathbf{k}}^{\eta}=0 \nonumber
\\(-\xi_N-E_{\mathbf{k}}^{\eta})v_{N\mathbf{k}}^{\eta}
+\sum_{M}F_{{\mathbf{k}}MN}^*u_{M\mathbf{k}}^{\eta}=0 \end{eqnarray} where
 $u_{N\mathbf{k}}^{\eta}$ is the coefficient of $\phi_{N\mathbf{k}}$ for the
Bogoliubov function $u^{\eta}_{{\mathbf{k}}}(\mathbf{r})$ and $v_{N{\mathbf{k}}}^{\eta}$ is the
coefficient of $\phi_{N-\mathbf{k}}^*$ for the function $v^{\eta}_{{\mathbf{k}}}(\mathbf{r})$,
and $\xi_{N} =(N+1/2 )\hbar \omega_c-\mu$. $\mu$ is the chemical potential. The off-diagonal elements
 $F_{{\mathbf{k}}NM}$ are:
 \begin{equation} F_{{\mathbf{k}}NM}=\int d{\mathbf{r}}\Delta({\mathbf{r}})\phi({\mathbf{r}})_{N,{\mathbf{k}}}^*
 \phi({\mathbf{r}})_{M,{\mathbf{-k}}}^* 
 \end{equation}
  and the order parameter is determined self-consistently as:
  \begin{equation} \Delta({\mathbf{r}})=g\sum_{{\mathbf{k}}\eta}u^{\eta}_{{\mathbf{k}}}({\mathbf{r}})
  v^{\eta}_{{\mathbf{k}}}({\mathbf{r}})^*(1-2f^{\eta}_{{\mathbf{k}}}) \end{equation}
  where $g$ is the coupling strength and $f^{\eta}_{{\mathbf{k}}}=(1+\exp (E_{\mathbf{k}}^{\eta}/k_BT))$ is the 
  fermi function. We neglect any finite Zeeman splitting for simplicity. Due to the translational symmetry the order
   parameter $\Delta({\mathbf{r}})$ is completely characterised by a finite set of parameters 
  $\Delta_j$~\cite{Big Mac1}. For notational simplicity we work in the lowest 
  Landau level approximation (LLL) (i.e $\Delta_{J\neq0}=0$) in which the center-of-mass motion of the Cooper-pairs 
  has the kinetic energy $\hbar\omega_c/2$ where $\omega_c$ is the cyclotron frequency. None of the conclusions
   in this paper are altered when this restriction is relaxed. When the chemical potential $\mu$ is at a Landau level 
 (i.e $n_f\equiv \mu/\hbar\omega_c-1/2=$ integer ) we have, in the normal state, exact degeneracy 
  between an electron state in the  Landau 
  level $n_f+m$ and a hole state in the  Landau level $n_f-m$. Likewise when the chemical potential is 
  exactly in between two Landau levels such that $n_f=n+1/2$ there is degeneracy between an electron
   in a level $n_f+m+1/2$ and a hole in a level $n_f-m-1/2$. When this is the case we expect the possible
  non-perturbative effects of a finite order-parameter to be strongest. We will show that the convergence radius
   for the Gor'kov 
  equations is indeed smallest when $n_f$ is an integer. To examine the validity of the Gor'kov expansion it 
  is convenient to use the following expression~\cite{Bardeen}
   for the difference $\Omega_S - \Omega_N$ in the thermodynamic potential between the mixed state and 
  the normal state:
  \begin{equation}\label{thermo} \Omega_S-\Omega_N=\frac{1}{g}\int d {\mathbf{r}} 
  |\Delta({\mathbf{r}})|^2-2k_BT
 \sum_{N{\mathbf{k}}} \ln(\cosh(\beta E_{N {\mathbf{k}}}/2))+2k_BTD\sum_N
 \ln(\cosh(\beta \xi_i/2))
 \end{equation}
 Here $D=\frac{VeB}{2\pi \hbar c}$ is the number of ${\mathbf{k}}$-vectors in the Brillouin
  zone,  $V$ is the volume, $B$ is the magnetic field, and $\beta=1/k_BT$. 
\section{Zero temperature}
 To illustrate the origin of the non-perturbative effect 
 for $T=0$ it is sufficient to examine the case when only one Landau level participates in the pairing and 
  $n_f$ is an integer. In this case the positive energy solution to equation\ (\ref{BdG}) is $E_{n_f {\mathbf{k}}}=
 |F_{{\mathbf{k}}n_fn_f}|$. Equation\ (\ref{thermo}) reduces to:
 \begin{equation} E_{gS}-E_{gN}=\frac{1}{g}\int d {\mathbf{r}}|\Delta({\mathbf{r}})|^2-\sum_
 {{\mathbf{k}}}E_{n_f{\mathbf{k}}}
 \end{equation}
 Since $|F_{{\mathbf{k}}n_fn_f}|\propto \Delta_0 \propto |\Delta({\mathbf{r}})|$ we see 
 that we obtain a linear term in $|\Delta({\mathbf{r}})|$ in equation\ (\ref{thermo}). This is a 
 non-perturbative term since the Gor'kov expansion only contains even powers of the order parameter. 
 This $T=0$ result is unaltered when we have many Landau levels participating in the pairing and it 
 agrees with the result obtained by Bahcall~\cite{Bahcall}. It is a trivial consequence of the fact that we 
 have to take the $T\rightarrow 0$ limit $k_BT\ln(2\cosh(\beta E_{N {\mathbf{k}}}\/2))\rightarrow
 E_{N {\mathbf{k}}}/2$ before we perturbatively expand the result in the size of the order parameter.
\section{Finite temperature}
\subsection{Quantum limit}
 For finite temperature the situation is different. It is now possible to expand 
 $\ln(2\cosh(\beta E_{N {\mathbf{k}}}/2))$ in powers of the order parameter and then check if 
 we obtain any non-perturbative terms, as proposed by Bahcall~\cite{Bahcall}.
 For notational simplicity we will again do the calculation in the quantum limit when only one Landau level 
participates in the 
 pairing. In section~\ref{several} we will treat the slight modifications in our result when more than one 
Landau level are 
 within the pairing width. The  quasiparticle energy is now $E_{{n\mathbf{k}}}=\sqrt{\xi_n^2+
  |F_{{\mathbf{k}}nn}|^2}$.
   We need to expand $ \ln(\cosh(\beta E_{{n\mathbf{k}}}/2))$ in 
  $|F_{{\mathbf{k}}nn}|^2$. Writing $\beta E_{{n\mathbf{k}}}/2=\sqrt{\epsilon^2+z^2}$ where 
$\epsilon\equiv\beta\xi_n/2$ 
  and $z=\beta|F_{{\mathbf{k}}nn}|/2$ we are lead to consider the analytic properties of the function 
 $\ln(\cosh(\sqrt{\epsilon^2+z^2}))$. The poles and branch cuts in the complex plane $z \in \mathcal{C}$ 
determine the convergence radius $r_0$ for a power series in $z$.  A simple analysis gives: 
 $r_0=\sqrt{\epsilon^2+\pi^2/4}$. The requirement 
 for the convergence of a perturbation series for $\ln(2\cosh(\beta E_{N {\mathbf{k}}}/2))$ is then 
\begin{equation}  \label{criterium} |F_{{\mathbf{k}}nn}| \leq \sqrt{\xi_n^2+\pi^2(k_BT)^2}\end{equation}
This requirement is most restrictive when the Landau level is at the chemical potential ($\xi_n=0$). We then 
have
  \begin{equation}  E_{{\mathbf{k}}}\leq k_BT\pi \end{equation} 
 Furthermore, we see that there will appear only even powers of $|F_{{\mathbf{k}}nn}|$ in the series. 
 This is true for general 
 $\mu$ (i.e also when $\xi_n=0$). So we have ruled out any non-perturbative cubic term in the expression 
 for $\Omega_S-\Omega_N$ thereby disproving earlier predictions based on a numerical analysis~\cite
 {Bahcall}. Doing the expansion and comparing with a standard expression for  the thermodynamic potential 
 based on Gor'kov's equations~\cite{Bruun} we find (not surprisingly) that it reproduces the Gor'kov series 
 term  by  term.  The convergence of the Gor'kov expansion is determined be equation (\ref{criterium}). It is  now clear 
 that the Gor'kov expansion is a high temperature series. So the 
 break down of the theory for $T=0$ is of no surprise. For finite $T$ we expect the Gor'kov series first become 
unreliable when there is a Landau level at the chemical potential. This is because the requirement in 
 equation (\ref{criterium})  is most restrictive when $\xi_n=0$ and because the superconductivity and thereby the 
 change in the quasiparticle energies (obtained by a self-consistent solution of equation\ (\ref{BdG}) is enhanced 
when there is a Landau level at the chemical  potential~\cite{Bruun}.
\subsection{Several Landau levels} \label{several}
 The above conclusions are essentially unaltered when there is more than one Landau level participating 
 in the pairing. We calculate the quasiparticle energies from equation\ (\ref{BdG}) perturbatively in 
 $F_{{\mathbf{k}}nm}$ using 
 degenerate and non-degenrate perturbation theory. Then we 
 expand $\ln(2\cosh(\beta E_{N {\mathbf{k}}}/2))$ in powers of the order parameter. The convergence radius 
for the series is again smallest when the chemical potential is at a Landau level. The only 
 complication is that we obtain both even and odd powers of $F_{{\mathbf{k}}nm}$ in the expression for the 
 quasiparticle energies. But the
 odd terms cancel in the expression for $\Omega_S-\Omega_N$ due to the fact that there are two 
 quasiparticle levels when $\xi_n \neq 0$ for which the odd powers in the expression for the energy have 
 opposite signs. There is only one positive energy solution for the case $\xi_{n_f}=0$ though. However, the odd
  terms from this solution vanish in the expression for  $\Omega_S-\Omega_N$ due to the fact that 
 $\partial^{2l+1}_x \ln(\cosh(x))|_{x=0}=0$ where $l$ is an integer. A long tedious calculation shows that we 
 recover the standard terms in the Gor'kov series. A sufficient condition for the convergence of the Gor'kov 
series  is
\begin{equation} \label{criterium2} E_{n{\mathbf{k}}}-\xi_n \leq \min \left[ 2k_BT\pi,k_BT
\sqrt{\beta^2\xi^2_n+\pi^2} \right]\end{equation} 
which has to hold for each quasiparticle level within the pairing region.
 So we expect  that the Gor'kov theory breaks down when significant portions of the quasiparticle bands lie 
 outside the regions defined in equation (\ref{criterium2}). It should be noted that one cannot ignore the contribution
  from higher quasiparticle levels to $\Omega_S-
 \Omega_N$ ($\xi_n \neq 0$). This  is easily seen from equation\ (\ref{thermo}) since 
 \begin{equation}  \ln(\cosh(\beta(\xi+\delta E)/2))-\ln(\cosh(\beta\xi/2))>\ln(\cosh(\beta(\delta E)/2))
 \ \ \ \xi,\delta E>0\end{equation}
 So any treatment which focuses only on the quasiparticle level at the chemical potential will ignore 
 important contributions to $\Omega_s$. Equation\ (\ref{criterium2}) can be transformed into the requirement:
 \begin{equation} <|\Delta({\mathbf{r}})|^2>\equiv\frac{1}{V}\int d{\mathbf{r}}|\Delta({\mathbf{r}})
 |^2 \leq 2\sqrt{\pi n_f}\pi^2(k_BT)^2\end{equation}
Based on an extensive numerical analysis, Norman \textit{et al}~\cite{Norman} have suggested a similar 
condition.
Since the Ginzburg-Landau equations are derived from the microscopic BCS theory using the Gor'kov 
expansion~{\cite{Parks}  it would be of interest to restate the above criterium in terms of Ginzburg-Landau
 parameters. Doing this we obtain:
\begin{equation}\frac{H_{c2}-H}{H_c(0)}\leq \sqrt{n_f}\pi^{1/2}(7\zeta(3))^{1/2}\beta_Ae^{\gamma}
(\kappa-\frac{1}{2\kappa})\left(\frac{T}{T_c}\right)^2
\end{equation}
 Here $\zeta(x)$ is Riemann's Zeta function, $\kappa$ is the Ginzburg Landau parameter, $\beta_A$ 
is the Abrikosov parameter, and $\gamma$ is Euler's constant. This restriction is always fulfilled 
for type II superconductors within the normal range of validity of the Ginzburg-Landau equations (i.e 
$|T-T_c|/T_c \ll 1$).

As an example of the breakdown of the perturbation series we have plotted the orderparameter 
$\Delta_0$ as a function of $n_f=\frac{\mu}{\hbar\omega_c}-1/2$. In figure\ 1 we have plotted both a numerical 
exact solution of the BdG-equations and the fourth order perturbative result using a method  
developed earlier~\cite{Bruun}.
 We have chosen the parameters such that   $\omega_d/\omega_c=5$, 
 $g/\hbar\omega_cl^2=8.2$ and $k_BT/\hbar\omega_c=0.3$ when $n_f=12$. As can be seen the perturbation theory agrees fairly well with the exact solution. 
The perturbative result tends to differ the most from the exact solution when the chemical potential is at a Landau
 level ($n_f$ integer). This is in agreement with the above remarks. To illustrate the temperature dependence of
 the convergence radius of the Gor'kov series we 
have in figure\ 2 again plotted $\Delta_0$ as a function of $n_f$ for a very low temperature.  As can be seen the
 perturbation series breaks down much earlier $\Delta_0\simeq 1500$ for 
this low temperature in agreement with equation (\ref{criterium2}). In figure\ 3 we have 
plotted the lowest quasiparticle level along a high symmetry direction
 in ${\mathbf{k}}$-space when $n_f=12$ for  the parameters used in figure\ 2. The horizontal line 
gives the  boundaries for $E_{{\mathbf{k}}}-\xi$ calculated from equation (\ref{criterium2}).  There are large parts of 
the quasiparticle band in ${\mathbf{k}}$-space lying outside the region of convergence of the perturbation 
expansion thereby  explaining the observed discrepancies between perturbation theory and 
the exact numerical result. 

\section{Conclusion}
In conclusion we have examined the recently debated validity of the Gor'kov expansion. The conclusion is 
 that although the degeneracy of the normal state levels gives large effects on the quasiparticle wavefunctions
 even for weak superconducting order these effects cancel in the expression for the thermodynamic potential
 and the Gor'kov expansion is correct for finite temperature. We have therefore ruled out the possibilty of a 
non-perturbative third order term in the expression for the thermodynamic potential. The range of validity of 
the Gor'kov expansion 
is given by equation\ (\ref{criterium2}) which shows that it is essentially a high temperature expansion. This 
requirement is always fulfilled within the range of validity of the Ginzburg-Landau equations leading to 
no inconsistencies. Furthermore we
 have the usual requirement $|E_{n{\mathbf{k}}}-\xi_n|\ll \hbar \omega_c$ for perturbation theory 
to work. We expect the Gor'kov expansion the break down first when the chemical potential is at a Landau 
level. Our results are confirmed by a comparison  between the results of an exact numerical solution to the 
BdG-equations and a Gor'kov series to fourth order in the order parameter.  
\\
\section{Acknowledgements}

The authors would like to acknowledge EPSRC grant GR/K 15619 (VNN) and The Danish Research Academy 
(GMB) for financial support.

\newpage
\begin{center}{Figure Captions}\end{center}
\bigskip
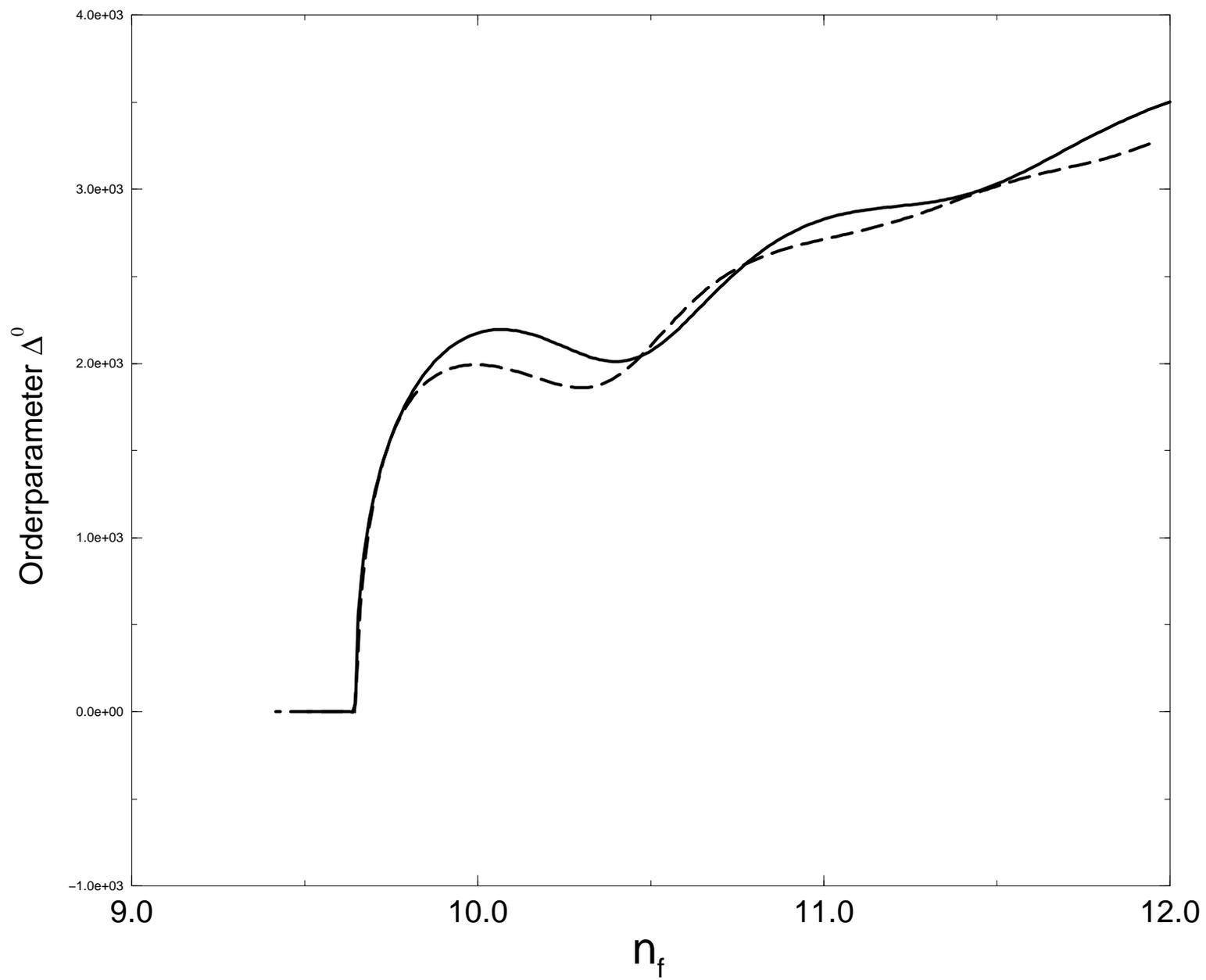
\noindent Fig.\ 1: The order parameter $\Delta _0$ vs $n_f=\mu/\hbar\omega_c -1/2$ calculated 
numerically  (solid line) and perturbatively to fourth order in $\Delta_0$ (dotted  line) for 
$\omega_d/\omega_c=5$, $\frac{g}{\hbar\omega_cl^2}=8.2$ and $k_BT/\hbar\omega_c=0.3$ 
when $n_f=12$.

\

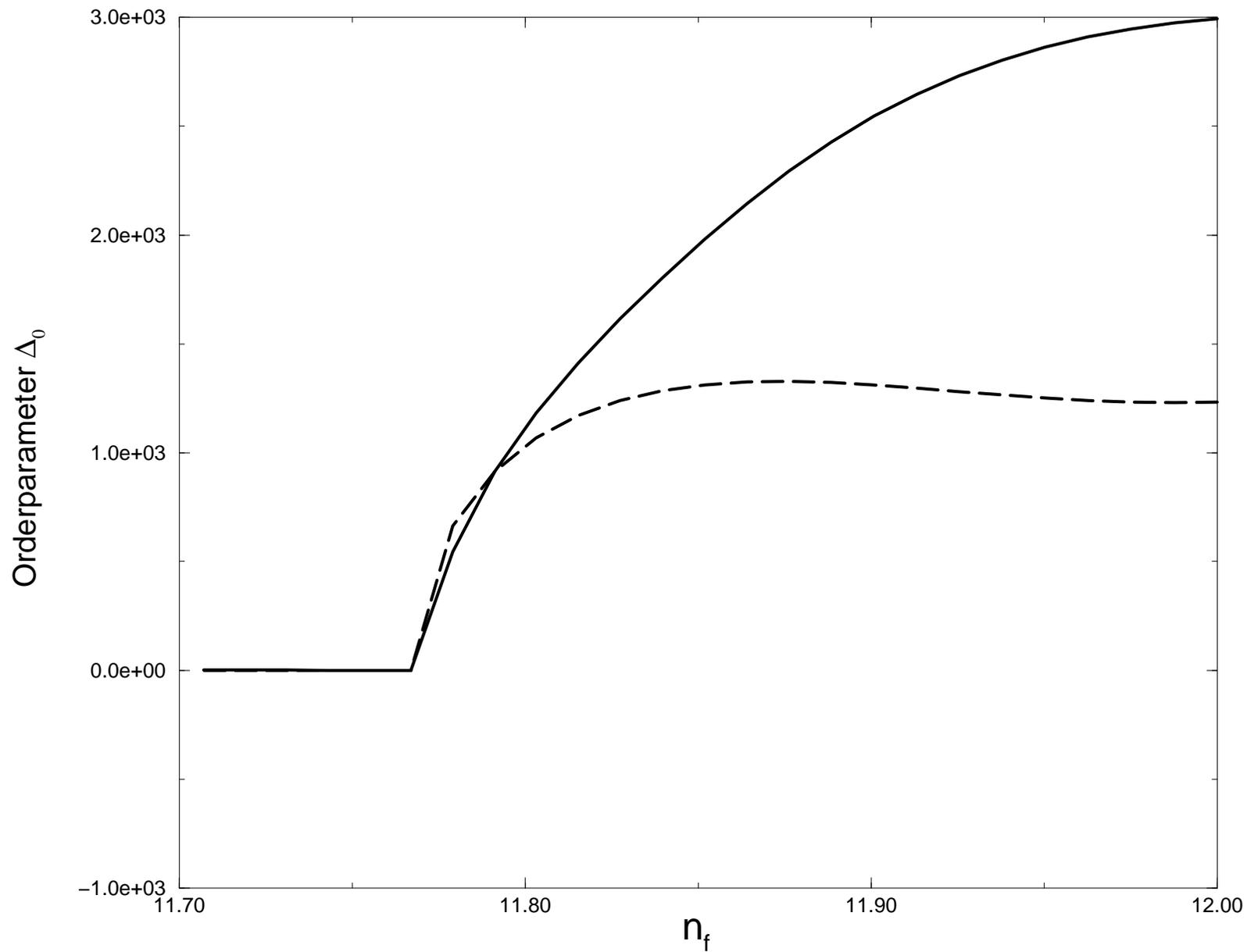
\noindent Fig.\ 2: The order parameter $\Delta _0$ vs $n_f=\mu/\hbar\omega_c -1/2$ calculated 
numerically  (solid line) and perturbatively to fourth order in $\Delta_0$ (dotted  line) for 
$\omega_d/\omega_c=5$, $\frac{g}{\hbar\omega_cl^2}=7.0$ and $k_BT/\hbar\omega_c=0.05$ 
when $n_f=12$.

\

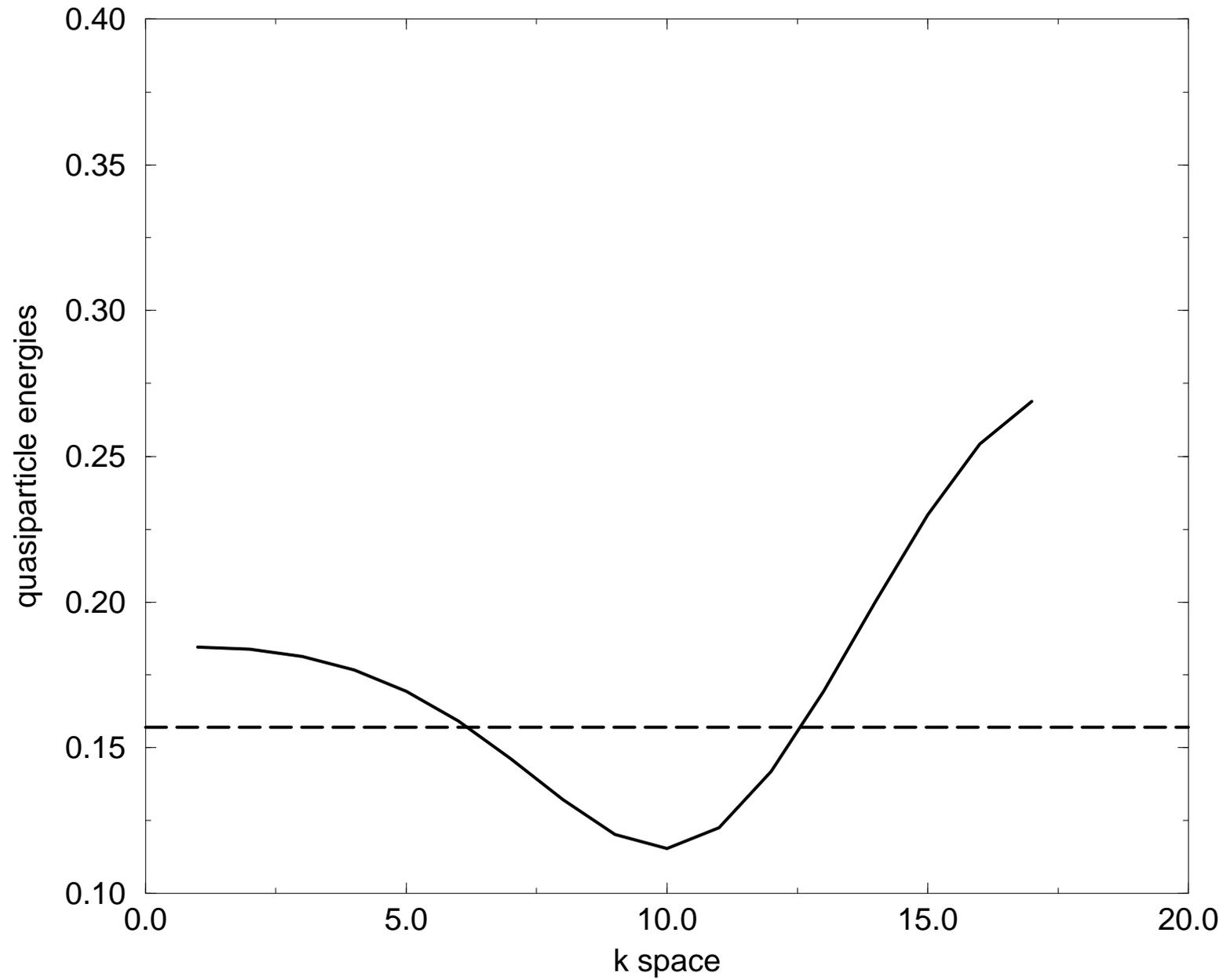
\noindent Fig.\ 3: The  lowest quasiparticle band in units of $\hbar \omega_c$ along a high 
symmetry line in \textbf{k}-space for 
$n_f=12$ . The dashed line marks the boundary defined by equation\ (\ref{criterium2}).

\end{document}